\documentclass[12pt]{article}

\begin{document} 

\title{
{\Large \bf Multi Parametric Deformed Heisenberg
Algebras: A Route to Complexity} 
}

\author{
 E. M. F. Curado and M. A. Rego-Monteiro \\
Centro Brasileiro de Pesquisas F\'\i sicas, \\ 
Rua Xavier Sigaud 150, 22290-180 - Rio de Janeiro, RJ, Brazil 
} 

\maketitle

\begin{abstract}
 
\indent

	We introduce a generalization of the Heisenberg algebra which
is written in terms of a functional of one generator of the algebra,
$f(J_0)$, that can be any analytical function. When $f$ is linear
with slope $\theta$, we show that the algebra in this case corresponds
to $q$-oscillators for $q^2 = \tan \theta$. The case where $f$ is
a polynomial of order $n$ in $J_0$ corresponds to a $n$-parameter
deformed Heisenberg algebra. The representations of the algebra,
when $f$ is any analytical function, are shown to be obtained
through the study of the stability of the fixed points of $f$ and
their composed functions. The case when $f$ is a quadratic polynomial
in $J_0$, the simplest non-linear scheme which is able to create
chaotic behavior, is analyzed in detail and special regions in the
parameter space give representations that cannot be continuously
deformed to representations of Heisenberg algebra.

\end{abstract}

\vspace{1cm}

\begin{tabbing}

\=xxxxxxxxxxxxxxxxxx\= \kill

non-linearity; chaos; Gauss number; 
q-analysis.
 \\

\end{tabbing}

\newpage

\section{Introduction}

Quantum algebras have first appeared in the algebraic Bethe ansatz
approach to quantum integrable one-dimensional models \cite{qism}. 
Since then, 
there have been several attempts to apply them in a broad range of 
physical phenomena \cite{zachos}. 

Associated to the omnipresent harmonic oscillator there is an
algebra known as Heisenberg algebra. The simple structure of
this algebra, that is described in terms of 
creation and annihilation
operators, and its particle interpretation has promoted it to
a paradigmatic tool in the second quantization approach.

A connection between these two topics appeared soon after 
the discovery of quantum algebras when it was 
found out that a generalization of Heisenberg algebra, known 
as q-oscillators, was necessary in order to realize $su_{q}(2)$ 
through the Jordan-Schwinger method \cite{qosc}. 

Guided, in part, by the wide range of physical applicability 
of Heisenberg
algebra there have been along the last ten years some effort 
in order to analyze possible physical relevance of $q$-oscillators
or deformed Heisenberg algebras \cite{varios}. 
The expected physical properties of
toy systems described by these generalized Heisenberg algebras
were analyzed and indications on how to solve an 
old puzzle in physics were obtained \cite{marco}. 

Recently, it was introduced an algebra, called logistic algebra,
that is a generalization of Heisenberg algebra where the 
eigenvalues 
of one generator of the algebra (the one that generalizes the
number operator) are given by functional iterations of the logistic
function. This algebra has finite- and infinite-dimensional
representations associated to the cycles of the logistic map and
infinite-dimensional representations related to the chaotic band
\cite{algebra1}, \cite{algebra2}.

A quantum solid Hamiltonian whose collective modes of vibration 
are described by oscillators satisfying the logistic algebra 
was constructed 
and it was analyzed the thermodynamic
properties of this model in the two-cycle and in 
a specific chaotic 
region of the logistic map. It is interesting to mention that
in the chaotic band this model shows a curious hybrid behavior
mixing classical and quantum behavior showing how a 
quantum system can present a non-standard quantum behavior 
\cite{algebra2}. 

In this paper, a generalization of the logistic algebra 
is constructed in such a way that the
eigenvalues of one generator is given by a functional iteration of 
a starting number. This functional could be any analytical 
function but, in order to study the properties of this algebra  
in detail, this function is taken as a polynomial of order $n$.
 
When the functional, $f(J_{0})$,  
is linear in $J_0$, where $J_0$ is the hermitean
generator of the algebra, i.e., $f(J_0) = r \, J_0 + s$, 
$r = q^2$ is shown to correspond to $q$-deformed Heisenberg
algebra or $q$-oscillators. The general case, $f(J_0) = \sum^n_{i=0}
r_i J_0^i$ is a $n$-parameter deformed Heisenberg algebra.
This algebra is, therefore,  a multi parametric 
deformation of Heisenberg algebra.

The representation theory is presented in detail for the linear
and quadratic cases since they are the paradigmatic ones. It is
shown that the essential tool in order to find the representations
of the algebra is the analysis of the stability of the fixed points
of the polynomial $f$ and their composed functions.

Related to the cycles of period 1, 2, 4, ... there are finite- and
infinite-dimensional representations of the algebra. The weights of
the finite-dimensional representations are given exactly by
the lowest values of the cycles. 

In the next section we present the general algebra and the general
representation theory. In section 3 we analyze the linear case,
their representations and its connection to $q$-oscillators. The
non-linear case or two-parameter deformed Heisenberg algebra is
presented in section 4 where it becomes evident the essential
role played by the analysis of the stability of the fixed points
of the polynomial $f$ and their composed functions in order to
obtain the finite- and infinite-dimensional representations of the
algebra. In section 5 we present our final comments and also introduce
a generalization of $su(2)$ in the sense discussed in 
this paper.

\section{Generalized Heisenberg algebra}

Let us consider an algebra generated by $J_{0}$, $J_{\pm}$ described
by the relations
\begin{eqnarray}
J_{0} \, J_{+} &=& J_{+} \, f(J_{0}) ,
\label{eq:alg1} \\
J_{-} \, J_{0} &=& f(J_{0}) \, J_{-} , 
\label{eq:alg2} \\
\left[ J_{+},J_{-} \right] &=& J_{0}-f(J_{0}) .
\label{eq:alg3}
\end{eqnarray}
Note that eq. (\ref{eq:alg2}) is the Hermitean conjugate of 
eq. (\ref{eq:alg1}), implying that 
$J_{-}=J_{+}^{\dagger}$ and $J_{0}^{\dagger}=J_{0}$, and $f(J_{0})$
is a general analytic function of $J_{0}$. The case where 
$f(J_{0})=r \, J_{0} \, (1-J_{0})$ was analyzed in refs. \cite{algebra1} 
and \cite{algebra2}. 
The above algebra relations are constructed in order that the eigenvalues
of operator $J_{0}$ are given by an iteration of an initial value
as will be clear in a moment.

Let us now show that the operator
\begin{equation}
C = J_{+} \, J_{-} - J_{0} = J_{-} \, J_{+} - f(J_{0}) , 
\label{eq:casimir}
\end{equation}
is a Casimir operator of the algebra.
Using the algebraic relations in eqs. (\ref{eq:alg1}-\ref{eq:alg3}) 
it is easy to see that
\begin{equation}
\left[ C,J_{0} \right] = \left[ C,J_{\pm} \right] = 0 , 
\label{eq:comute}
\end{equation}
i.e., $C$ is one Casimir operator of the algebra.

We start now analyzing the representation theory of the algebra when 
the function $f(J_{0})$ is a general analytic function of  
$J_{0}$. 
In this section we obtain the general equations for an $n$-dimensional
representation and in the next sections we solve these equations for
linear and quadratic polynomials $f(J_{0})$ finding out the finite- 
and infinite-dimensional representations for the linear and quadratic
cases that are the paradigmatic ones.

We assume we have an $n$-dimensional irreducible representation
of the algebra given in eqs. (\ref{eq:alg1}-\ref{eq:alg3}). 
The hermitean operator $J_{0}$
can be diagonalized. Consider the state $|0\rangle$ with the lowest 
eigenvalue of $J_{0}$
\begin{equation}
J_{0} \, |0\rangle = \alpha_{0} \, |0\rangle .
\label{eq:alfa0}
\end{equation}
For each value of $\alpha_{0}$ and the parameters of the algebra
we have a different vacuum that for simplicity will be denoted by
$|0\rangle$. Moreover, it will be clear in the next sections, when we shall
solve the representation theory for the linear and quadratic
polynomials $f(J_{0})$, that the allowed values of $\alpha_{0}$
depend on the parameters of the algebra. 

Let $| m \rangle$ be a normalized eigenstate of $J_{0}$, 

\begin{equation}
    J_{0} |m \rangle = \alpha_{m} |m \rangle \, . 
    \label{eq:alfam}
\end{equation}
Applying eq. (\ref{eq:alg1}) on $|m \rangle$ we have  

\begin{equation}
    J_{0} (J_{+} |m \rangle) = J_{+} f(J_{0}) |m \rangle = 
    f(\alpha_{m}) (J_{+} |m \rangle ) \, .
    \label{eq:j+}
\end{equation} 
Thus, we see that $J_{+} |m \rangle $ is a $J_{0}$ 
eigenvector with eigenvalue $f(\alpha_{m})$.  
Starting from $|0 \rangle$ and applying successively $J_{+}$ on 
$|0 \rangle$ we create different states with $J_{0}$ eigenvalue 
given by 
\begin{equation}
    J_{0} \left( J_{+}^m |0 \rangle \right) = 
    f^m (\alpha_{0}) \left( J_{+}^m |0 \rangle \right) \, ,
    \label{eq:j+m}
\end{equation}
where $f^m (\alpha_{0})$ denotes the $m$-th iterate of $f$.  Since 
the application of $J_{+}$ creates a new vector, whose respective 
$J_{0}$ eigenvalue has iterations of $\alpha_{0}$ through $f$ 
increased by one unit, it is 
convenient to define the new vectors $J_{+}^m |0 \rangle$ as 
proportional to $|m \rangle$ and we then call $J_{+}$ a raising 
operator.  Note that  
\begin{equation}
\alpha_m = f^m(\alpha_0) = f(\alpha_{m-1}) \, ,
    \label{eq:alfam3}
\end{equation}
where $m$ denotes the number of iterations of $\alpha_{0}$ 
through $f$.  

Following the same procedure for $J_{-}$,  applying eq. (\ref{eq:alg2}) 
on $|m+1 \rangle$, we have 
\begin{equation}
    J_{-} J_{0} |m+1 \rangle = f(J_{0}) \left( J_{-} |m+1 \rangle \right) =
    \alpha_{m+1} \left( J_{-} |m+1 \rangle \right) \, .
    \label{eq:alfam2}
\end{equation}
This shows that $J_{-} |m+1 \rangle$ is also a $J_{0}$ eigenvector with 
eigenvalue $\alpha_{m}$.  Then, $J_{-} |m+1 \rangle $ is proportional 
to $|m \rangle$ showing that $J_{-}$ is a lowering operator.  

Since we consider 
$\alpha_{0}$ the lowest $J_{0}$ eigenvalue, we must have
\begin{equation}
J_{-} \, |0\rangle = 0 .
\label{eq:vacuum}
\end{equation}
As was shown in \cite{algebra2}, depending on the function $f$ 
and its initial value $\alpha_{0}$, it may happen that the 
$J_{0}$ eigenvalue of state $|m+1 \rangle$ is lower than the one 
of state $|m \rangle$.   Thus, as we exemplify  
in section IV of this paper, 
given an arbitrary analytical function $f$ 
(and its associated algebra in eqs. (\ref{eq:alg1}-\ref{eq:alg3})) 
in order to satisfy eq. (\ref{eq:vacuum}),   
the allowed values of $\alpha_{0}$ are chosen in such a way that the  
iterations $f^m (\alpha_{0})$ ($m \geq 1$) are 
always bigger than $\alpha_{0}$. 
Then, eq.(\ref{eq:vacuum}) must be checked for every  
function $f$, giving consistent vacua for specific values 
of $\alpha_{0}$. This analysis is made in sections 
3 and 4 where we find the parameter regions with   
consistent representations. 

In general we obtain
\begin{eqnarray}
J_{0} \, |m-1\rangle &=& f^{m-1}(\alpha_0) \, |m-1\rangle , \; \; \; m = 1,2, 
\cdots \; , 
\label{eq:b1} \\
J_{+} \, |m-1\rangle &=& N_{m-1} \, |m\rangle , 
\label{eq:b2} \\
J_{-} \, |m\rangle &=& N_{m-1} \, |m-1\rangle ,
\label{eq:b3}
\end{eqnarray}
where $N_{m-1}^2 = f^{m}(\alpha_0)-\alpha_0$. 
We observe that if we put $m=0$ in eq. (\ref{eq:b3}) then 
$N_{-1}$ is equal to zero, which is consistent with eq. (\ref{eq:vacuum}).
Eqs. (\ref{eq:b1}-\ref{eq:b3}) are easily
proven by induction. In order to verify 
eqs. (\ref{eq:b1}-\ref{eq:b3}) for $m = 1$, apply
eq. (\ref{eq:alg1}) on the state vector $|0\rangle$ obtaining 
$J_0 \, (J_+ |0\rangle) = f(\alpha_0) \, (J_+ |0\rangle)$. Thus, we define 
$|1\rangle \equiv \frac{1}{N_0} J_+ |0\rangle$ where $N_0$ is a constant to be
determined. It is easy to see that $J_0 |1\rangle = f(\alpha_0) |1\rangle$. The
constant $N_0$ can be determined by imposing that the state vector
$|1\rangle$ has unit norm and with the use of eq. (\ref{eq:alg3}), 
we get $N_0^2 = f(\alpha_0)-\alpha_0$. As
the last step of this check apply eq. (\ref{eq:alg3}) on the state $|0\rangle$.
Using eqs. (\ref{eq:alfa0}) and (\ref{eq:vacuum}) we get 
$J_- |1\rangle = N_0 |0\rangle$. 
Then, eqs. (\ref{eq:b1}-\ref{eq:b3}) 
are verified for $m = 1$.

Now, suppose eqs. (\ref{eq:b1}-\ref{eq:b3}) are valid for m. 
Apply $J_0$ on eq. (\ref{eq:b2}) and
use eq. (\ref{eq:alg1}) on the left hand side, this gives
\begin{equation}
J_{0} \, |m\rangle = f^{m}(\alpha_0) \, |m\rangle .
\label{eq:j0fm}
\end{equation} 
Applying eq. (\ref{eq:alg1}) on the state $|m\rangle$ and using 
eq. (\ref{eq:j0fm}) we are
allowed to suppose that there exists a state vector $|m+1\rangle$ such that
\begin{equation}
|m+1\rangle \, = \, \frac{1}{C(m)} \, J_+ |m\rangle  \, ,
\label{eq:proporcional}
\end{equation}
where $C(m)$ is a constant. This constant is determined by imposing 
that the state vector $|m+1\rangle$ has unit norm
\begin{eqnarray}
1 &=& \langle m+1|m+1\rangle = \, \frac{1}{C(m)^2} \, \langle m|J_- \, J_+ |m\rangle =  \nonumber \\
  &=& \frac{1}{C(m)^2} \, \left[ \langle m|J_+ \, J_- |m\rangle \, + \, 
\langle m|(-J_0 \, + \, f(J_0)) |m\rangle \right] =  \nonumber  \\
  &=& \frac{1}{C(m)^2} \, \left( N_{m-1}^2 - f^m(\alpha_0) + 
f^{m+1}(\alpha_0) \right) \, , 
\label{eq:unitnorm}
\end{eqnarray}
which gives $C(m)^2 = N_m^2 = f^{m+1}(\alpha_0) - \alpha_0$.

Applying eq. (\ref{eq:alg2}) on $|m\rangle$ and using eqs. (\ref{eq:b1}-\ref{eq:b3}) 
plus the value of $N_{m}$ we obtain the
last equation we wanted. Putting everything together we recover
eqs. (\ref{eq:b1}-\ref{eq:b3}) for $m \mapsto m+1$ and the proof is complete.

Note that eqs. (\ref{eq:b1}-\ref{eq:b3}) define a general $n$-dimensional 
representation for the algebra in eqs. (\ref{eq:alg1}-\ref{eq:alg3}). In order to
solve it, i.e., to construct the conditions under which we have 
finite- and infinite-dimensional representations we have to specify 
the functional $f(J_0)$. It is easy to see that if we choose 
$f(J_0) = J_0 + 1$ the algebra given by eqs. (\ref{eq:alg1}-\ref{eq:alg3}) 
becomes with this choice the Heisenberg algebra 
for $A$, $A^\dagger$ and $N=A^\dagger A$ where $A = J_{-}$, 
$A^\dagger = J_{+}$ and $N = J_{0}$. Note that the Casimir 
operator in eq. (\ref{eq:casimir}), 
that in the general case has eigenvalue equal to $-\alpha_{0}$, 
becomes in this case 
$C = A^\dagger A - N$ which is identically null.   
We shall see in the next section
that the choice $f(J_0) = r \, J_0 + s$ corresponds to a 
one-parameter deformed Heisenberg algebra and if we take a
functional with linear and quadratic terms (besides a constant term) 
we have  a quadratic Heisenberg algebra 
or a two-parameter deformed Heisenberg algebra that
will be analyzed in section 4.

Another very interesting observation is that, as mentioned in the
beginning of this section, the algebraic relations eqs. (\ref{eq:alg1}) 
and (\ref{eq:alg2})
are constructed in such a way that the eigenvalues of operator
$J_0$ are iterations of an initial value $\alpha_0$ through the
function $f$ as shown in eq. (\ref{eq:b1}). Then, the increasing complexity
of function $f$  will correspond to an increasing complex behavior 
of the eigenvalues of $J_0$ \cite{yorke}. In fact, as already shown in 
refs. (\cite{algebra1}, \cite{algebra2}) 
choosing the logistic map for $f$ it could give rise to a 
chaotic behavior 
of the eigenvalue of $J_0$. Moreover, as will be clear in the next
sections, it is this iteration aspect of the algebra that will 
allow us to find their representations through the analysis of the
stability of the fixed points of the function $f$ and their
composed functions.

\section{The linear case}

In this section we are going to find the representations for the
algebra defined by the relations given in eqs. (\ref{eq:alg1}-\ref{eq:alg3}) 
considering
$f(J_0) = r \, J_0 + s$. The algebra relations can be rewritten
for this case as
\begin{eqnarray}
\left[ J_{0},J_{+} \right]_{r} &=& s \, J_{+} , 
\label{eq:coml1} \\
\left[ J_{0},J_{-} \right]_{r^{-1}} &=& -\frac{s}{r} J_{-} , 
\label{eq:coml2} \\
\left[ J_{+},J_{-} \right] &=& (1-r) \, J_{0}-s ,
\label{eq:coml3}
\end{eqnarray}
where $\left[ a,b \right]_{r} \equiv a \, b - r \, b \, a$ is the
$r$-deformed commutation of two operators $a$ and $b$.

It is very simple to realize that, for $r=1$ and $s$ arbitrary,
the above algebra is the Heisenberg algebra for $A$, $A^{\dagger}$
and $N$ where $A=J_-/\sqrt{s}$, $A^{\dagger}=J_+/\sqrt{s}$ and
$N=J_0/s$. In this case the Casimir operator given in 
eq. (\ref{eq:casimir}) is
null. Then, for general $r$ and $s$ the algebra defined in 
eqs. (\ref{eq:coml1}-\ref{eq:coml3})
is a one-parameter deformed Heisenberg algebra and 
generally speaking the algebra
given in eqs. (\ref{eq:alg1}-\ref{eq:alg3}) is a generalization of the
Heisenberg algebra. 

It is easy to see for the general linear case that
\begin{eqnarray}
f^m(\alpha_0) &=& r^m \, \alpha_0 + s \, (r^{m-1}+r^{m-2}+
\cdots +1) 
\label{eq:gauss1}\\
&=& r^m \alpha_0 + s \, \frac{r^m -1}{r-1} ,  \nonumber
\end{eqnarray}
thus,
\begin{equation}
N_{m-1}^2 = f^m(\alpha_0)-\alpha_0 = \left[ m \right]_r \, N_{0}^2
\label{eq:gauss2}
\end{equation}
where $\left[ m \right]_r \equiv (r^m -1)/(r-1)$ is the Gauss
number of $m$ and $N_0^2 = \alpha_0 \, (r-1)+s$.

Let us search for finite-dimensional representations of the 
linear Heisenberg algebra. Our approach is the following:
we start from the vacuum state $|0\rangle$ and apply repeatedly the
operator $J_+$ arriving, for specific values of $\alpha_0$,
$r$ and $s$, eventually to 
$J_+ |n-1\rangle = 0$ for a $n$-dimensional
representation. From eq. (\ref{eq:b2}) we see that the set of parameters 
providing an $m$-dimensional representation, using eq. (\ref{eq:gauss2}), 
is computed from 
\begin{eqnarray}
N_0^2 & = & \alpha_0 \, (r-1) + s \; > \; 0 \; , \nonumber \\
N_1^2 & =& \left[ 2 \right]_r \, N_0^2 \; > \; 0 \; , \nonumber \\
& & \cdot \;\; \cdot \;\; \cdot 
\label{eq:nm} \\
N_{m-2}^2 & =& \left[ m-1 \right]_r \, N_0^2 \; > \; 0 \; , \nonumber \\
N_{m-1}^2 & =& \left[ m \right]_r \, N_0^2 = 0 \; . \nonumber 
\end{eqnarray}
The solutions for $\left[ m \right]_r = 0$ are given by 
$r = exp(2 \pi ik/m)$ for $k=1,2,\cdots ,m-1$, ($k=0$ corresponds
to Heisenberg algebra that we are not considering at the moment) but
since $J_0$ is taken hermitean, the only interesting finite 
dimensional solution is a two-dimensional ($m=2$) 
representation with $r=-1$ and $s > 2 \alpha_0$. There is of course 
a trivial one-dimensional representation where the weight
of the representation is the fixed point 
$\alpha_{0} = \alpha^{*} = s/(1-r)$ and 
$r \in (-1,1) \cup (1,\infty)$. We have also a marginal uninteresting
one-dimensional solution obtained
for $r \rightarrow \infty$ and $s/r^2 = $finite.

The infinite-dimensional solutions are more interesting. In this
case we must solve the following set of equations:
\begin{equation}
N_m^2 \; > \; 0 \, , \; \; \; \forall m, \; m = 0,1,2, \cdots \;\; . 
\label{eq:nm2}
\end{equation}
Apart from the Heisenberg algebra given by $r=1$, the solutions
are
\begin{eqnarray}
\mbox{type} \; \mbox{I} &:& \; r > 1 \;\; \mbox{and} \;\; \alpha_0 > 
\frac{s}{1-r} \;\;\;\; 
\mbox{or} 
\label{eq:tipo}\\
\mbox{type} \; \mbox{II} &:& \; -1 < r < 1 \;\; \mbox{and} \;\; \alpha_0 < 
\frac{s}{1-r} \;\; ,
\nonumber
\end{eqnarray}
with matrix representations
\begin{equation}
    J_{0}= \left(  
    \begin{array}{ccccc}
        \alpha_{0} & 0 & 0 & 0 & \ldots  \\
        0 & \alpha_{1} & 0 & 0 & \ldots  \\
        0 & 0 & \alpha_{2} & 0 & \ldots  \\
        0 & 0 & 0 & \alpha_{3} & \ldots  \\
        \vdots & \vdots & \vdots & \vdots & \ddots 
    \end{array}
    \right)  , \hspace{0.1cm}
       J_{+}= \left(  
    \begin{array}{ccccc}
        0 & 0 & 0 & 0 & \ldots  \\
        N_{0} & 0 & 0 & 0 & \ldots  \\
        0 & N_{1} & 0 & 0 & \ldots  \\
        0 & 0 & N_{2}^ø & 0 & \ldots  \\
        \vdots & \vdots & \vdots & \vdots & \ddots 
    \end{array}
    \right) , \hspace{0.1cm}
    J_{-} = J_{+}^{\dag} \hspace{0.1cm} .
    \label{eq:matriz}
\end{equation}
Note that for type I solutions the eigenvalues of $J_0$ , as can be 
easily computed
from eqs. (\ref{eq:b1})  and (\ref{eq:alfam3}), go to infinite as 
we consider eigenvectors 
$|m\rangle$ 
with increasing value of $m$. Instead, for type II solutions the
eigenvalues go to the value $s/(1-r)$, the fixed point of $f$,
as the state $|m\rangle$ increase.

The reason for this asymptotic behavior of the eigenvalues of $J_0$
is simple. It is clear from eqs. (\ref{eq:b1})  and 
(\ref{eq:alfam3}) that the 
eigenvalues of $J_0$
are given by the functional iteration of 
$f(\alpha) = r \, \alpha + s$ for the starting number $\alpha_0$. 
Moreover, the stability of the 
fixed point of $f(\alpha)$ is directly related to the asymptotic 
behavior of the eigenvalue of $J_0$. If the fixed point of 
$f(\alpha)$ is stable ($-1 < r < 1$) or unstable ($r > 1$)
the eigenvalues of $J_0$ go to the fixed point 
$\alpha^{\star} = s/(1-r)$ or to infinite respectively 
since they are 
given by iterations of $\alpha_0$ through the function $f$.
Finally, we mention that the allowed values of $\alpha_0$ in
eq. (\ref{eq:tipo}) are purely algebraic conditions that comes 
from our choice that the representations of the algebra have 
always a lowest-weight vector.

The interesting and certainly \emph{unexpected} connection we have just 
analyzed between the infinite-dimensional representations of the
linear Heisenberg algebra and the classification of the different
types of fixed point and their stability 
will become more relevant in the next section where we shall
consider the quadratic case $f(J_0) = q \, J_0^2 + r \, J_0 +s$.
In this case, even the finite dimensional representations will be 
connected to the fixed point analysis through the attractors of $f$.

It is interesting to note that in eq. (\ref{eq:gauss2}) we obtained, considering
the linear case, the well-known Gauss number of $m$  as
\begin{equation}
\frac{N_{m-1}^2}{N_0^2} = \frac{r^m - 1}{r-1} = \left[ m \right]_r \; .
\label{eq:gauss3}
\end{equation}

It is possible to look at the above equation the other way round
and to define a general Gauss number $\left[ m \right]_{general}$ 
for the case of arbitrary $f$ as
\begin{equation}
\left[ m \right]_{general} \equiv \frac{N_{m-1}^2}{N_0^2} = 
\frac{f^m(x)-x}{f(x)-x} \;\; .
\label{eq:gauss4}
\end{equation}
Of course, this definition gives
\begin{eqnarray}
\left[ m \right]_{general} \; & \longrightarrow & \; m \;\;\;\;\;\;\; 
for \;\;\; f(x) = x + s \;\;\; , 
\label{eq:gauss5}\\
\left[ m \right]_{general} \; & \longrightarrow & \; \left[ m \right]_r 
\;\;\; for \;\;\; f(x) = r \, x + s \;\;\; . \nonumber
\end{eqnarray}

Finally, it is easy to see that there is a direct relation between
the linear Heisenberg algebra given in 
eqs. (\ref{eq:coml1}-\ref{eq:coml3}) and the standard
$q$-oscillators. In fact, defining
\begin{eqnarray}
J_0 & = & q^{2 N} \, \alpha_0 + s \, \left[ N \right]_{q^2} \;\; , 
\label{eq:qosc1} \\
\frac{J_+}{N_0} & = & a^{\dagger} \,  q^{N/2} \;\; ,
\label{eq:qosc2}\\
\frac{J_-}{N_0} & = & q^{N/2} \, a \;\; , 
\label{eq:qosc3}
\end{eqnarray}
we see that $a$, $a^{\dagger}$ and $N$ satisfy the usual 
$q$-oscillator relations \cite{qosc}
\begin{eqnarray}
&&a \, a^{\dagger} - q \, a^{\dagger} \, a = q^{-N} \;\; , \;\;
a \, a^{\dagger} - q^{-1} \, a^{\dagger} \, a = q^{N} \;\; , 
\label{eq:qosc4}\\
&&\left[ N,a \right] = -a \;\; , \;\;
\left[ N,a^{\dagger} \right] = a^{\dagger} \;\; . \nonumber 
\end{eqnarray}
Note that, Heisenberg algebra is obtained from 
(\ref{eq:qosc1}-\ref{eq:qosc3}) for
$q \rightarrow 1$ and $\alpha_0 = 0$.

\section{The non-linear case}

In this section we consider the algebra defined by 
eqs. (\ref{eq:alg1}-\ref{eq:alg3}) 
for $f(x) = t \, x^2 + r \, x + s$ .  In this case the algebra becomes 
\begin{eqnarray}
\left[ J_0,J_+ \right]_r &=& t \, J_+ \, J_0^2 + s \, J_+ \;\;\; ,
\label{eq:comq1}\\
\left[ J_0,J_- \right]_{r^{-1}} &=& -\frac{t}{r} \, J_0^2 \, J_- - 
\frac{s}{r} \, J_- \;\;\; ,
\label{eq:comq2}\\
\left[ J_+,J_- \right] &=& -t \, J_0^2 + (1-r) \, J_0 - s \;\;\; .
\label{eq:comq3}
\end{eqnarray}
Of course, for $t=0$ we
recover the linear (or $r$-deformed) Heisenberg algebra given
in eqs. (\ref{eq:coml1}-\ref{eq:coml3}) and for $t=0$ and $r=1$ 
the standard Heisenberg algebra.

We focus now on the analysis of 
eqs. (\ref{eq:alfa0},\ref{eq:vacuum}-\ref{eq:b3}), 
aiming to find
the finite- and infinite-dimensional representations of the
above quadratic Heisenberg algebra. Following an observation
done at the end of the previous section we shall find the 
algebra representations through the analysis and the stability
of the fixed points of $f(x) = t \, x^2 + r \, x + s$ 
and their composed functions.

One clear way to do this is to perform a graphical analysis
of the function $f$. Let us graph $y=f(x)$ together with
$y=x$. Where the lines intersect we have $x=y=f(x)$, so that
the intersections are precisely the fixed points. Now, for a
point $x_0$, different from the fixed point, in order to follow
its path through iterations with the function $f$ we perform
the following steps
\begin{enumerate}
\item
move vertically to the graph of $f(x)$,
\item
move horizontally to the graph of $y=x$, and
\item
repeat steps 1, 2, etc. (in figure 1  
it is shown the example of 
the Heisenberg algebra, where $f(J_{0}) = J_{0} + 1$) .
\end{enumerate}

There are three cases to be analyzed: (I) $\Delta < 0$, 
(II) $\Delta = 0$ and (III) $\Delta > 0$, for 
$\Delta = (r-1)^2 - 4\,t\,s$. In the first case there is no
fixed point and it is easy to see by a graphical analysis that
only $t > 0$ corresponds to infinite-dimensional representations
($N_m^2 \not= 0$, $\forall m$, $m \in Z^+$) having 
\emph{lowest} weight states as desired (see figure 2(a)). 
Then, case (I) provides infinite-dimensional representations 
with lowest weight $\alpha_0$ for the
value of the parameters
\begin{equation}
t > 0 \;\;\; , \;\;\; (r-1)^2 - 4 \, t \, s < 0 \;\;\; \mbox{and} \;\;\;
\alpha_0 \in \Re \;\;\; .
\label{eq:t+01}
\end{equation}

In case (II), $t > 0$ as well and we have one fixed point given by
$\alpha^{\star} = (1-r)/2t$. This fixed point corresponds to a 
trivial one-dimensional representation of the algebra for 
$\alpha_0 = \alpha^{\star}$ since $N_0 = 0$. Besides this trivial
one-dimensional representation we have for case (II) 
infinite-dimensional representations 
with lowest weight $\alpha_0$ for the set of parameters
(see figure 2(b))

 \begin{equation}
 t > 0 \;\;\; , \;\;\; (r-1)^2 - 4 \, t \, s = 0 \;\;\; \mbox{and} \;\;\;
 \alpha_0 \in \Re, \, \alpha_{0} \neq (1-r)/2q  \;\;\; .
 \label{eq:t+02}
 \end{equation}
Case (III) is less trivial. In this case it is also possible to have 
attractors of period 1, 2, 4, $\cdots$ and even a chaotic region
in the space of parameters ($t$, $r$, $s$, $\alpha_0$). Thus, there
are regions in this space associated to finite- and 
infinite-dimensional representations. In what follows, we analyze
completely the cases of attractors of period 1, 2 and give an example
of the chaotic behavior of the algebra. For shortness, the  
analysis from now on will be done only for $t>0$; the $t<0$ 
behaviour is similar, with no conceptually significant difference.

We recall that a fixed point $\alpha^{\star}$, where by definition
$\alpha^{\star}$ is solution of the equation 
$\alpha^{\star} = f(\alpha^{\star})$, is stable if 
$|f^{'}(\alpha^{\star})|$ is smaller than one and is unstable if
it is greater than one. For case (III) the fixed points are
\begin{equation}
\alpha^{\star}_{\pm} = \frac{1-r \pm \sqrt{\Delta}}{2 \, t} \;\;\; .
\label{eq:alfastar}
\end{equation}
The fixed point $\alpha_+^{\star}$ is always unstable and computing
the derivative of $f$ at $\alpha_-^{\star}$ we have that 
$\alpha_-^{\star}$ is stable for a set of $t$, $r$ and $s$ such that
$0 < \Delta < 4$ (we stress again that this analysis 
is for $t>0$). 
For this set of ($t$, $r$, $s$) we must search
for the region of $\alpha_0$ that corresponds to lowest-weight
states. It is easy to realize that the region 
$\alpha^{\star}_- < \alpha_0 < \alpha^{\star}_+$ has to be eliminated
since it does not correspond to a representation with lowest-weight
state, i.e., there will always exist an $n > 0$ such that 
$\alpha_{n} < \alpha_{0}$ if 
$\alpha^{\star}_{-} < \alpha_{0} < \alpha^{\star}_{+}$. 

For the allowed values of $\alpha_0$ corresponding to 
infinite-dimensional representations with lowest-weight state, i.e.,
$- \infty < \alpha_0 < \alpha^{\star}_-$ and 
$\alpha_0 > \alpha^{\star}_+$, there are two types of asymptotic 
behaviors for the eigenvalues of $J_0$. They can go to infinite or 
go to the fixed point $\alpha^{\star}_-$. In order to identify
these two regions consider the point $f(\alpha^{\star}_+)$. There
is another point, denominated $\alpha^m$, that gives 
$f(\alpha^{\star}_+)$, i.e., $f(\alpha^m) = f(\alpha^{\star}_{+}) = 
\alpha^{\star}_{+}$, this point is given by 
\begin{equation}
\alpha^m = \frac{-1-r-\sqrt{\Delta}}{2 \, t} \;\;\; .
\label{eq:alfamq}
\end{equation}
It is easy to verify that the set of ($t$, $r$, $s$, $\alpha_0$)
such that
\begin{equation}
0 < \Delta < 4 \;\; \mbox{and} \;\; 
\left\{ \begin{array}{ll}
\mbox{(a)} & -\infty < \alpha_0 < \alpha^m \;\; \mbox{or} \;\; 
\alpha^{\star}_+ < \alpha_0 < \infty \\
\mbox{(b)} & \alpha^m < \alpha_0 < \alpha^{\star}_-
\end{array}
\right.
\label{eq:delta1}
\end{equation}
corresponds to infinite-dimensional representations where the
asymptotic eigenvalues of $J_0$ in case (a) go to infinite and
in case (b) go to the asymptotic value $\alpha^{\star}_-$,    
see figure 2(c). 
Moreover, $\Delta > 0$ and 
$\alpha_0 = \alpha^{\star}_-$ or $\alpha_0 = \alpha^{\star}_{+}$ 
correspond to the trivial finite one-dimensional representation. 
Note that in case (b), eq. (\ref{eq:delta1}), future 
iterations of $\alpha_{0}$ (that are 
always bigger than $\alpha_{0}$) will not increase monotonically. 
This is a specific example were a non-monotonic function $f$ presents 
a non-monotonic behavior of iterations of $\alpha_{0}$, with a consistent 
vacuum $|0 \rangle$.

Next step is to consider the set of parameters 
($t$, $r$, $s$, $\alpha_0$) such that the function $f(\alpha) = 
t \, \alpha^2 + r \, \alpha + s$ has an attractor of period 2.
This will permit us to find infinite-dimensional representations
where the asymptotic behavior of the eigenvalues of $J_0$ is
infinity or an attractor of period 2. Moreover, when the weight
of the representation is the lowest value of the  
attractor there will be a set of
parameters ($t$, $r$, $s$) corresponding to a 2-dimensional
representation.

In order to perform that analysis we must study the fixed points
of $f^2(\beta) \equiv f(f(\beta))$, i.e., the points $\beta^{\star}$
satisfying $\beta^{\star} = f^2(\beta^{\star})$ that are different
from the previous one-cycle (attractors of period 1). They are
\begin{equation}
\beta^{\star}_{\pm} = \frac{-1-r \pm \sqrt{\Delta_1}}{2 \, t} \;\;\; ,
\label{eq:beta1}
\end{equation}
where $\Delta_1 = -3 - 2 \, r + r^2 - 4 \, t \, s$. Since the
fixed points of $f^2$, $\beta^{\star}_{\pm}$, have the same tangent it is 
sufficient to analyze the stabilization region for one of them.
It is simple to see that this region is given by the set ($t$, $r$, $s$)
such that $4 < \Delta < 6$. We see that for $\Delta = 4$ the
one-cycle solution looses stability and starts the stabilization
region for the two-cycle solution. Then, the set of 
($t$, $r$, $s$, $\alpha_0$) such that
\begin{equation}
4 < \Delta < 6 \;\; \mbox{and} \;\; 
\left\{ \begin{array}{ll}
\mbox{(c)} & -\infty < \alpha_0 < \alpha^m \;\; \mbox{or} \;\; 
\alpha^{\star}_+ < \alpha_0 < \infty  \; ,\\
\mbox{(d)} & \alpha^m < \alpha_0 < \beta^{\star}_- \; ,
\end{array}
\right.
\label{eq:delta2}
\end{equation}
corresponds to infinite-dimensional representations where the
asymptotic eigenvalues of $J_0$ in case (c) go to infinite and
in (d) go to the lowest value of the stable two-cycle
attractor with values $\beta^{\star}_{\pm}$. 

In this case there is also a set of parameters, for $\Delta > 4$, 
corresponding to a
2-dimensional representation. Note that if we take the weight of 
the representation as
\begin{equation}
\alpha_0 = 
\beta^{\star}_- = \frac{-1-r- \sqrt{\Delta_1}}{2 \, t} \;\;\; ,
\label{eq:beta2}
\end{equation}
we have a two-dimensional representation with matrix representation
given by
\begin{equation}
    J_{0}= \left(  
    \begin{array}{cc}
        \beta_-^{\star} & 0 \\
        0  & \beta_+^{\star}   
    \end{array}
    \right)  , \hspace{0.1cm}
       J_{+}= \left(  
    \begin{array}{cc}
        0 & 0  \\
        N_{0} & 0  
    \end{array}
    \right) , \hspace{0.1cm}
    J_{-} = J_{+}^{\dag} \hspace{0.1cm} ,
    \label{eq:matriz2}
\end{equation}
where $N_0$ is computed for $\Delta > 4$ and $\alpha_0$
given in eq. (\ref{eq:beta2}).

Clearly, for $\Delta > 6$, we will have other cycles, of 
length 4, 8, \ldots, $2^k$ \ldots,  entering then in the chaotic region  
and displaying, in the region ($\alpha_{m}$, $\alpha_{+}^{\star}$), 
exactly the same scenario the logistic map shows. To give 
an example of the chaotic region one chooses a point in the parameter 
space presenting two chaotic bands. 
This point corresponds to the numeric values $t=1$, $r=2$ and 
$s= -1.543591$, see figure 3. 
Actually, there is a whole surface in the parameter space $(t, r, s)$, 
in which this point is included,  exhibiting  
these two chaotic bands. Clearly also, chaos implies infinite-dimension 
representation and, for the example above, 
the eigenvalues of $J_{0}$ belong, mainly, to the $\alpha$-region 
limited by the two chaotic bands showed in figure 3.  
The frequency of a specific eigenvalue 
is given by the relative height of the band at this value. If we call 
the lowest value of $\alpha$ of the  
two bands by $\alpha_{chaos}^m$, the allowed range for 
the lowest weight values of possible representations  
in this example is 
$\alpha_{0} \, \, \in \, \, (\alpha^m,\,  \alpha_{chaos}^m$). 

In the case where $t<0$ the whole region outside the interval 
$(\alpha^m, \, \alpha_{+}^{\star})$ is not allowed, contrary to the case 
$t>0$.  The lowest fixed point is always unstable, also contrary to 
the case of positive values of $t$, where the highest fixed point was 
always unstable.  But the general sequence of attractors and chaotic 
regions is exactly the same as is well-known. A study of a particular 
case of $t<0$, the logistic case, was done in \cite{algebra1, algebra2}.

\section{Final comments}

In this paper we have presented the first steps towards the
complete analysis of the algebra described by the relations in
eqs. (\ref{eq:alg1}-\ref{eq:alg3}). This algebra can be rewritten for the polynomial
$f(J_0) = \sum^{n}_{i=0} a_i J_0^{i} $ as 
\begin{eqnarray}
\left[ J_0,J_+ \right]_{a_1} &=& a_0 J_+ + \sum^{n}_{i=2} 
a_i J_+ J_0^{i} \;\;\; ,
\label{eq:comg1}\\
\left[ J_0,J_- \right]_{a_1^{-1}} &=& -\frac{a_0}{a_1}  J_- - 
\sum^{n}_{i=2} \frac{a_i}{a_1} J_0^i J_- \;\;\; , 
\label{eq:comg2}\\
\left[ J_+,J_- \right] &=& -\sum^{n}_{i=2} a_i J_0^i + 
(1-a_1) \, J_0 - a_0 \;\;\; .
\label{eq:comg3}
\end{eqnarray}
The linear case, $f(J_0) = a_0 + a_1 \, J_0$, corresponds to
Heisenberg algebra for $a_1 = 1$ and to $a_1^2$-deformed 
Heisenberg algebra otherwise. The representation theory was 
shown to be directly related to the stability analysis of the
fixed point of the function $f$ and their composed functions.

The linear and quadratic cases were analyzed in detail. The 
finite-dimensional representations correspond to lowest-weight
being the lowest value of the attractors  
of period 1, 2, 4, \ldots . Moreover, associated
to each attractor there is a parameter region providing an
infinite-dimensional representation. We expect that this relation
between representations and stability analysis of the fixed
points of $f$ and their composed functions will be the same for
any analytical function $f$. In fact, in higher-order 
polynomials there will be the possibility to have, simultaneously, 
more than one attractor, each one with its own basin of attraction 
in the parameter space.  In spite of this, inside one particular 
basin of attraction the scenario is the same as analysed here 
in the non-linear case.

It is interesting to mention that there are parameter regions 
corresponding to certain representations that cannot be smoothly
deformed to a representation of Heisenberg algebra. An obvious
example is the so-called Logistic algebra where 
$f(J_0) = r \, J_0 (1-J_0)$ is chosen as the logistic map for $J_0$.
It is clear that this algebra cannot be deformed to Heisenberg
algebra even if it is a generalization of it in the sense
discussed in this paper.

Last, but not least, we have the feeling that the approach
we have presented in this paper be, in a certain sense, universal.
In this approach we construct the non-linear 
generalization of a given 
undeformed algebra and its representation theory is directly
related to the classification of the fixed points - and their 
stability - of a function $f$ (and their composed functions) that 
generates the algebra.  

In fact, it is possible to construct another iterative algebra
as
\begin{eqnarray}
J_{0} \, J_{-} &=& J_{-} \, f(J_{0}) ,
\label{eq:coms1}\\
J_{+} \, J_{0} &=& f(J_{0}) \, J_{+} ,
\label{eq:coms2}\\
\left[ J_{+},J_{-} \right] &=& J_{0} (J_0 + 1)-
f(J_{0}) (f(J_{0}) + 1) \;\;\; ,
\label{eq:coms3}
\end{eqnarray}
with Casimir
\begin{equation}
C = J_{+} \, J_{-} + f(J_{0}) (f(J_{0}) + 1) = 
J_{-} \, J_{+} + J_{0} (J_0 + 1)  \;\;\; , 
\label{eq:casimirs}
\end{equation}
where $J_- = J_+^{\dagger}$, $J_0^{\dagger} = J_0$ and $f(J_0)$
is an analytical function in $J_0$. Note that if $f(J_0)$ is the
simplest linear functional $f(J_0) = J_0 -1$ we obtain the relations
and the Casimir of the $su(2)$ algebra. It is tempting to investigate,
as we did in this paper for the iterative algebra in 
eqs. (\ref{eq:alg1}-\ref{eq:alg3}),
the above algebra for more complicated functionals $f(J_0)$.

\vspace{0.7 cm}

\noindent
{\bf Acknowledgments:} The authors thank a member of the Editorial 
Board for valuable comments. We also thank PRONEX and CNPq (Brazil) 
for partial support.

\newpage

\centerline{{\large\bf Figure Captions}}

\vskip 2\baselineskip
\noindent
{\bf Fig. 1:} \qquad Iterations of $\alpha_{0}$ for the Heisenberg 
algebra. The eigenvalues $\alpha_{n}$ increase by a constant factor 
as $n$ increases.

\vskip \baselineskip
\noindent
{\bf Fig. 2(a):} \qquad Iterations of $\alpha_{0}$ for the case I: 
\, $\Delta < 0$.  As it is easily seen, $\alpha_{n}$ goes to 
infinity as $n \rightarrow \infty$. This figure was plotted for  
the values $t=1$, $r=-1.5$ and $s=2.5$.

\vskip \baselineskip
\noindent
{\bf Fig. 2(b):} \qquad Iterations of $\alpha_{0}$ for the case II:
\, $\Delta = 0$.  Also in this case, for $\alpha \neq \alpha^{\star}$,  
$\alpha_{n}$ goes to 
infinity as $n \rightarrow \infty$. This figure was plotted for  
the values $t=1$, $r=-2$ and $s=9/4$.

\vskip \baselineskip
\noindent
{\bf Fig. 2(c):} \qquad Iterations of $\alpha_{0}$ for the case III:
\, $ 0 < \Delta < 4$. \,    
$\alpha_{0}^a$ is a starting point belonging to the regions 
$\alpha_{0} < \alpha^m$ or $\alpha_0 > \alpha^{\star}_+$, whose 
future iterations tend to infinity; \,  $\alpha_{0}^b$ is a 
starting point belonging to 
the region $\alpha^m < \alpha_0 < \alpha^{\star}_-$, and whose 
future iterations tend to the fixed point $\alpha_{-}^{\star}$. 
This figure was plotted for  
the values $t=0.8$,\,  $r=-4$ and $s=6$.

\vskip \baselineskip
\noindent
{\bf Fig. 3:} \qquad Histogram of the chaotic bands corresponding 
to the points $t=1$, \, $r=$ and $s=-1.543591$.

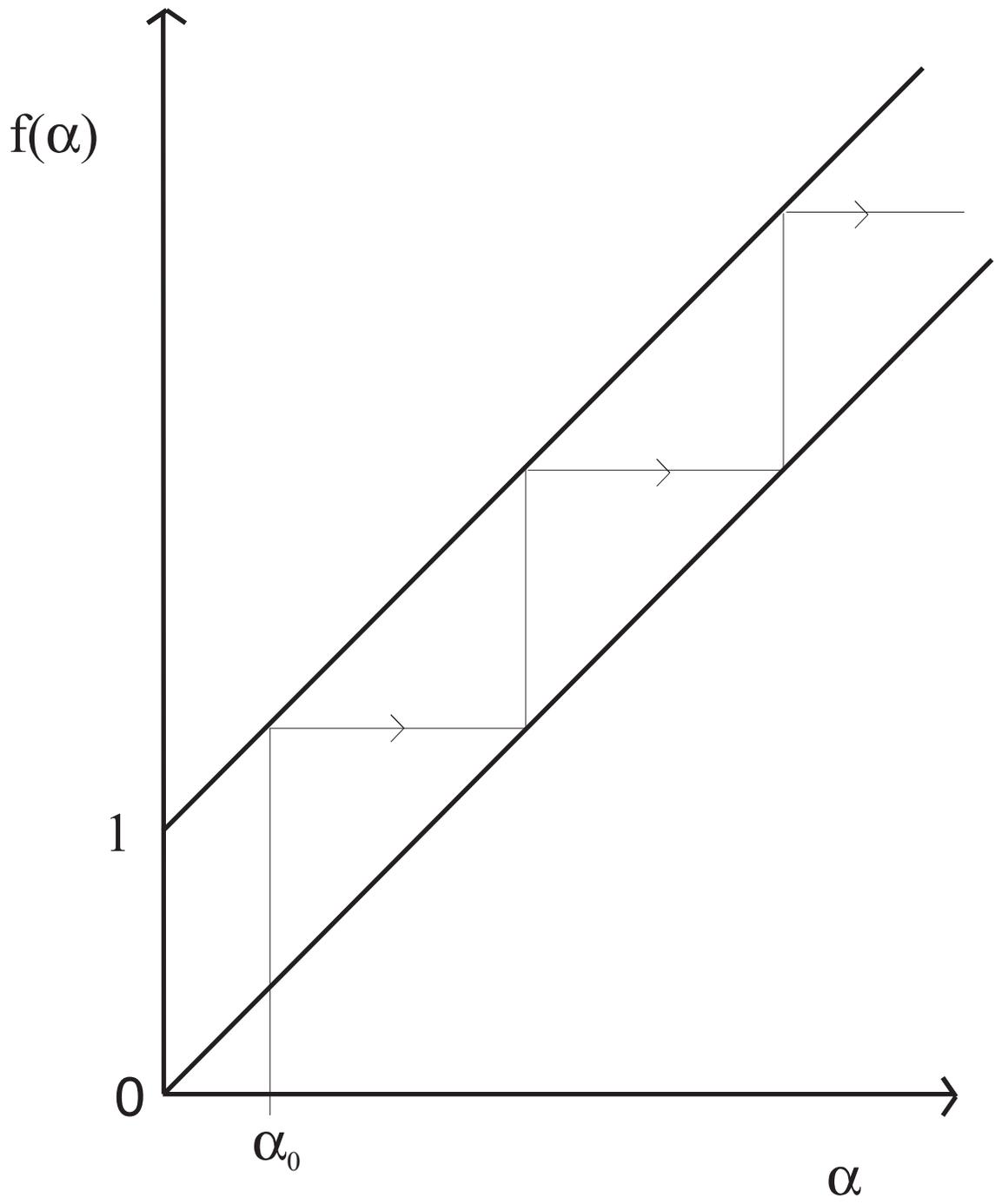
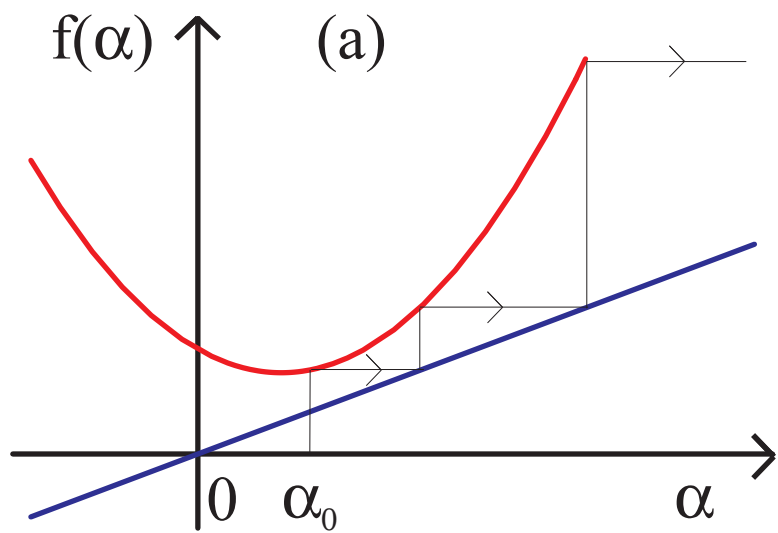
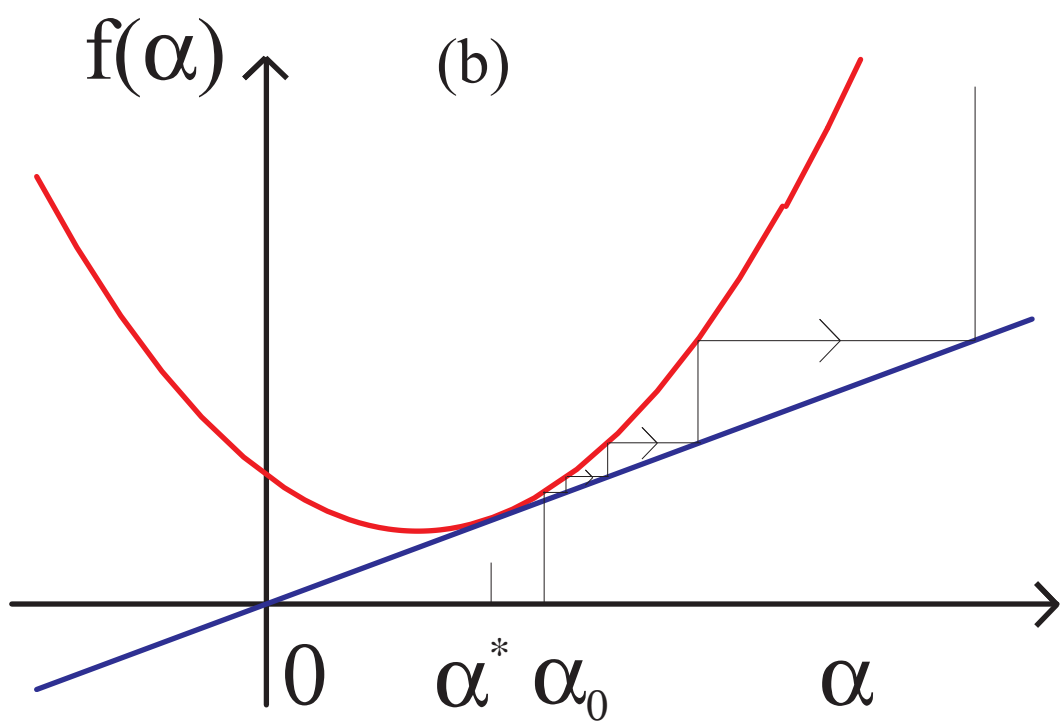
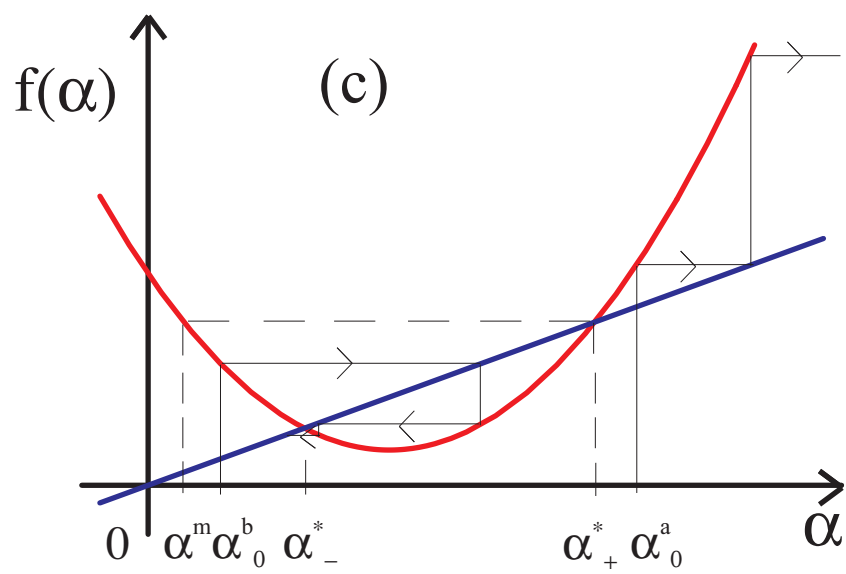
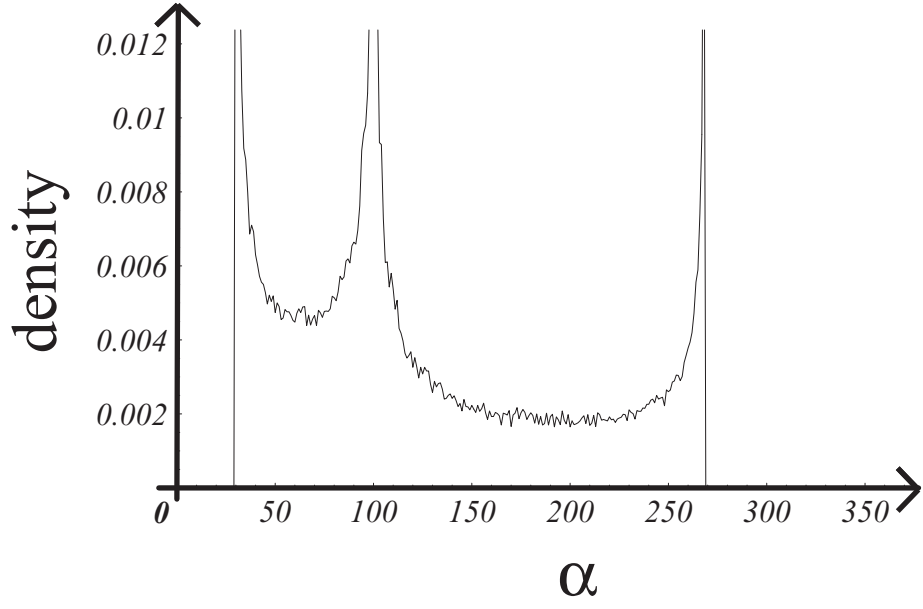
\end{document}